\input{aipcheck}

\documentclass[
    ,final            
  ]
  {aipproc}

\layoutstyle{6x9}

\begin{document}

\title{Effective Action and Schwinger Pair Production in Strong QED}


\author{Sang Pyo Kim}{
  address={Department of Physics, Kunsan National University,
Kunsan 573-701, Korea}, altaddress={Asia Pacific Center for
Theoretical Physics, Pohang 790-784, Korea}}

\begin{abstract}
Some field theoretical aspects, such as the effective action and
Schwinger pair production, are critically reviewed in strong QED.
The difference of the boundary conditions on the solutions of the
field equation is discussed to result in the effective action both
in the Coulomb and time-dependent gauge. Finally, the apparent
spin-statistics inversion is also discussed, where the WKB action
for bosons (fermions) works well for fermion (boson) pair-production
rate.
\end{abstract}

\maketitle


\section{Introduction}

The development of QED has laid a cornerstone for quantum field
theories, whose renormalization scheme has been regarded as a
paradigm for quantum field theories. However, in spite of enormous
successes of QED, such as prediction and measurement of the
anomalous magnetic moment, it has been studied and tested only in
the weak-field limit. QED processes are classified into weak and
strong QED according to the strength of electromagnetic field
involved. The critical strength is $E_c = m^2 c^3/\hbar q~(1.3
\times 10^{16}~{\rm V/cm})$ for the electric field (electron) and
$B_c = m^2 c^2/\hbar q ~(4.4 \times 10^{13}~{\rm G})$ for the
magnetic field. With an external field applied, the free Maxwell
theory, ${\cal L} = - {\cal F} = - F_{\mu \nu} F^{\mu \nu}/4$, gets
quantum corrections due to interactions with virtual charged pairs
from the vacuum, which lead to the effective action ${\cal L}_{\rm
eff} = - {\cal F} + \delta {\cal L} ({\cal F}, {\cal G})$, where
${\cal G} = F^*_{\mu \nu} F^{\mu \nu}/4$. The prominent consequence
of strong QED is that the imaginary part of $\delta {\cal L}$ for a
strong electric field leads to the vacuum decay and the real part
plays the role of a media for light propagation.

Historically, Sauter, Heisenberg and Euler, and Weisskopf studied as
early as thirties the effective action of strong electromagnetic
fields \cite{Sauter} and Schwinger used the proper-time method to
derive the exact one-loop effective action in a gauge invariant form
in a constant electromagnetic field \cite{Schwinger}. Though the
original motivation was a theoretical interest in quantum field
theory, the past two decades have witnessed enormous attractions in
strong QED. It is partly because the X-ray free electron laser from
Linac Coherent Light Source (LCLS) at SLAC \cite{Ringwald} and
Extreme Light Infrastructure (ELI) \cite{Dunne-Gies} will produce an
electric field near the critical strength for electron-position pair
production. These terrestrial experiments are expected to provide
many interesting physics. More astonishing physics comes from
astrophysical objects, neutron stars. Neutron stars have magnetic
fields ranging from $10^{8}~{\rm G}$ to $10^{15}~{\rm G}$ and more
than one-tenth of them are believed to have magnetic fields stronger
than $10^{14}~{\rm G}$, the so-called magnetars
\cite{Wood-Thompson}, at least one order greater than the critical
strength.

To find the effective action or to compute the pair-production rate
is a nontrivial task for a generic profile of electromagnetic field,
and only certain field configurations have been worked out exactly
(for review, see Ref. \cite{Dunne}). The Sauter-type electric field
\cite{Sauter32} that acts effectively for a finite period, $E (t) =
E_0 {\rm sech}^2 (t/T)$, or is confined to a local region, $E (x) =
E_0 {\rm sech}^2 (x/L)$, has been used to calculate the
pair-production rate by employing approximate schemes
\cite{Kim-Page02,Kim-Page06,Kim-Page07,GK,Dunne-Schubert,DWGS,KRX},
whose results are compared with the exact result \cite{Nikishov}.
The resolvent technique \cite{Dunne-Hall} and the evolution operator
method together with a gamma function renormalization scheme
\cite{Kim-Lee08,KLY} have been used to find the effective action for
the Sauter-type electric field, $E (t) = E_0 {\rm sech}^2 (t/T)$.
The purpose of this proceedings is to clarify some of issues of the
effective action and Schwinger pair production. First, the boundary
conditions on the solutions of field equation will be discussed for
the time-dependent and the Coulomb gauges. The Coulomb gauge, which
is a convenient choice for an electric field distributed in space,
requires a different boundary condition for the effective action
from the time-dependent gauge. Second, an interesting observation is
made that there seems to be an apparent spin-statistic inversion for
the WKB action for the pair-production rate, where the WKB action
for bosons works well for fermion pair production and vice versa.

\section{Effective Action in Coulomb Gauge}

One may find the pair-production rate either from the solutions of
the Klein-Gordon and Dirac equation or the effective action. At the
one-loop level, the solutions of the field equation carry all the
information about the effective action and thereby the
pair-production rate. The physical mechanism for Schwinger pair
production by an electric field in the time-dependent gauge is quite
similar to particle production by a time-changing spacetime: the
ingoing wave at one asymptotic region is scattered by either the
electric field or the spacetime metric and splits into a mixture of
positive and negative frequency waves at the other asymptotic region
\cite{DeWitt}. So, to find the effective action, one has to impose
the boundary condition that initially there is only an ingoing wave
which is scattered by interactions. With this boundary condition
imposed, the effective action is determined by the Bogoliubov
coefficient for the outgoing waves. Using the evolution operator and
regularizing gamma function, the exact one-loop effective action was
obtained for a constant electric field and the Sauter-type electric
field in the time-dependent gauge \cite{Kim-Lee08,KLY}.

On the other hand, the field equation for a spatially localized
electric field in the Coulomb gauge becomes a tunneling problem. To
simplify the physics, let us consider an electric field along the
$x$-direction in the Coulomb gauge. The Klein-Gordon equation for
scalar QED and the Dirac equation for spinor QED have the (spin
diagonal) Fourier component [in natural units with $\hbar = c = 1$]
\begin{eqnarray}
\Bigl[\partial_x^2 + (\omega - q A_0 (x))^2 - ( m^2 + {\bf
k}_{\perp}^2 + 2 i \sigma q E (x)) \Bigr] \phi_{\omega {\bf
k}_{\perp} \sigma} (x) = 0, \label{time-comp}
\end{eqnarray}
where $\sigma = 0$ for spin-0 bosons and $\sigma = \pm 1/2$ for
spin-1/2 fermions. The equation describes the tunneling problem
under the inverted harmonic potential. The incident wave function
corresponds to the ingoing solution and the reflected wave function
to the outgoing solution \cite{Kim-Page02}:
\begin{eqnarray}
\phi_{\omega {\bf k}_{\perp} \sigma} (x) = \alpha_{\omega {\bf
k}_{\perp} \sigma} \varphi_{\omega {\bf k}_{\perp} \sigma} (x) +
\beta_{\omega {\bf k}_{\perp} \sigma} \varphi^*_{\omega {\bf
k}_{\perp} \sigma} (x).
\end{eqnarray}
Here, $\varphi_{\omega {\bf k}_{\perp} \sigma}$ is the ingoing wave
function. The tunneling wave function
\begin{eqnarray}
\phi_{\omega {\bf k}_{\perp} \sigma} (x) = \gamma_{\omega {\bf
k}_{\perp} \sigma} \varphi_{\omega {\bf k}_{\perp} \sigma} (x)
\end{eqnarray}
corresponds to pair production.

From the causality with respect to the group velocity, the flux
conservation law reads
\begin{eqnarray}
|\alpha_{\omega {\bf k}_{\perp} \sigma}|^2 = |\beta_{\omega {\bf
k}_{\perp} \sigma}|^2 - (-1)^{2 |\sigma| + 1} |\gamma_{\omega {\bf
k}_{\perp} \sigma}|^2,
\end{eqnarray}
whose relative ratios lead to the Bogoliubov relation
\cite{Kim-Page02,Kim-Page05}
\begin{eqnarray}
\Bigl| \frac{\alpha_{\omega {\bf k}_{\perp} \sigma}}{\beta_{\omega
{\bf k}_{\perp} \sigma}} \Bigr|^2 + (-1)^{2 |\sigma| + 1} \Bigl|
\frac{\gamma_{\omega {\bf k}_{\perp} \sigma}}{\beta_{\omega {\bf
k}_{\perp} \sigma}} \Bigr|^2 = 1.
\end{eqnarray}
As the tunneling probability is related with pair production
 and the reflection probability with
vacuum-to-vacuum transition \cite{Kim-Page02}, one may identify the
Bogoliubov coefficients as
\begin{eqnarray}
\mu_{\omega {\bf k}_{\perp} \sigma} = \frac{\alpha_{\omega {\bf
k}_{\perp} \sigma}}{\beta_{\omega {\bf k}_{\perp} \sigma}}, \quad
\nu_{\omega {\bf k}_{\perp} \sigma} = \frac{\gamma_{\omega {\bf
k}_{\perp} \sigma}}{\beta_{\omega {\bf k}_{\perp} \sigma}}.
\end{eqnarray}
Now, the Bogoliubov transformation may take the form
\cite{Kim-Page08}
\begin{eqnarray}
\hat{a}^{\rm out}_{\omega {\bf k}_{\perp} \sigma} = \mu_{\omega {\bf
k}_{\perp} \sigma} \hat{a}^{\rm in}_{\omega {\bf k}_{\perp} \sigma}
+ \nu^*_{\omega {\bf k}_{\perp} \sigma} \hat{b}^{{\rm
in}\dagger}_{\omega {\bf k}_{\perp} \sigma},
\end{eqnarray}
where $\hat{a}$ and $\hat{b}$ denote particle and anti-particle
operators, respectively. Then, the outgoing vacuum is
\begin{eqnarray}
\vert 0, {\rm out} \rangle = \prod_{\omega {\bf k}_{\perp} \sigma}
U_{\omega {\bf k}_{\perp} \sigma} \vert 0, {\rm in} \rangle,
\end{eqnarray}
where $U_{\omega {\bf k}_{\perp} \sigma}$ is the evolution operator
in the Coulomb gauge given by
\begin{eqnarray}
U_{\omega {\bf k}_{\perp} \sigma} = e^{\xi_{\omega {\bf k}_{\perp}
\sigma} \hat{a}^{{\rm in} \dagger}_{\omega {\bf k}_{\perp} \sigma}
\hat{b}^{{\rm in} \dagger}_{\omega {\bf k}_{\perp} \sigma}}
e^{\theta_{\omega {\bf k}_{\perp} \sigma} (\hat{a}^{{\rm in}
\dagger}_{\omega {\bf k}_{\perp} \sigma} \hat{a}^{\rm in}_{\omega
{\bf k}_{\perp} \sigma} + \hat{b}^{{\rm in} \dagger}_{\omega {\bf
k}_{\perp} \sigma} \hat{b}^{\rm in}_{\omega {\bf k}_{\perp} \sigma}
- (-1)^{2 |\sigma| + 1})} e^{- \xi^*_{\omega {\bf k}_{\perp} \sigma}
\hat{a}^{\rm in}_{\omega {\bf k}_{\perp} \sigma} \hat{b}^{\rm
in}_{\omega {\bf k}_{\perp} \sigma}}. \label{u-op}
\end{eqnarray}
Here, the parameters $\xi_{\omega {\bf k}_{\perp} \sigma}$ and
$\theta_{\omega {\bf k}_{\perp} \sigma}$ are determined by the
Bogoliubov coefficients only. The last factor of Eq. (\ref{u-op})
annihilates pairs and the middle factor gives only a phase factor
but the prefactor creates pairs. Thus, the outgoing vacuum always
contains pairs of particles and anti-particles with the opposite
momenta, implying pair production. Then, it follows that mean number
of produced pairs for a given quantum number is
\begin{eqnarray}
{\cal N}_{\omega {\bf k}_{\perp} \sigma} =  |\nu_{\omega {\bf
k}_{\perp} \sigma}|^2 = \Bigl| \frac{\gamma_{\omega {\bf k}_{\perp}
\sigma}}{\alpha_{\omega {\bf k}_{\perp} \sigma}} \Bigr|^2.
\end{eqnarray}
The one-loop effective action given by \cite{Kim-Lee08,KLY}
\begin{eqnarray}
{\cal L}^{\rm eff} =  - (-1)^{2 |\sigma| + 1} i \sum_{\omega {\bf
k}_{\perp} \sigma} \ln (\mu^*_{\omega {\bf k}_{\perp} \sigma}),
\label{eff act}
\end{eqnarray}
leads to the the general relation between the imaginary part of the
effective action and the mean number of produced pairs:
\begin{eqnarray}
2 ({\rm Im} {\cal L}^{\rm eff}) = - (-1)^{2 |\sigma| + 1} \sum_{\omega {\bf k}_{\perp}
\sigma} \ln [ 1 - (1)^{2 |\sigma| + 1} {\cal N}_{\omega {\bf
k}_{\perp} \sigma}].
\end{eqnarray}

For the sake of simplicity, let us consider a constant electric
field in the Coulomb gauge, $A_0 = - Ex$. Now, the tunneling
solution at the asymptotic region at $x = - \infty$ is
\begin{eqnarray}
D_p (\zeta) = \frac{\sqrt{2 \pi}}{\Gamma(-p)} e^{-i (p+1)\pi/2} D_{-
p-1} ( i \zeta) + e^{-i p \pi} D_{p} (- \zeta),
\end{eqnarray}
while at the asymptotic region $x = + \infty$ it is
\begin{eqnarray}
\phi_{\omega {\bf k}_{\perp} \sigma} (x) = D_p(\zeta),
\end{eqnarray}
where $D_p$ is the parabolic cylinder function and
\begin{eqnarray}
\zeta = \sqrt{\frac{2}{qE}} e^{i \pi/4} (\omega + qE x), \quad p^* =
- \frac{1}{2} - i \frac{m^2 + {\bf k}_{\perp}^2 + 2 i \sigma qE}{2
(qE)}.
\end{eqnarray}
Thus, the Bogoliubov coefficient is found to be
\begin{eqnarray}
\mu_{\bf k} = \frac{\sqrt{2 \pi}}{\Gamma(-p)} e^{i (p-1)\pi/2}.
\end{eqnarray}
There is a caveat in finding the effective action from Eq. (\ref{eff
act}). Now, the argument of the gamma function is the complex
conjugate of that of the time-dependent gauge in Refs.
\cite{Kim-Lee08,KLY}. So, the contour integral for the proper-time
integral using the gamma function regularization in Refs.
\cite{Kim-Lee08,KLY} should be taken in the fourth quadrant instead
of the first quadrant. The renormalized exact one-loop effective
action is then given by
\begin{eqnarray}
{\cal L}^{\rm eff} &=& (-1)^{2 |\sigma| + 1} \frac{(2 |\sigma| +
1)}{16 \pi^2} {\cal P} \int_0^{\infty} \frac{ds}{s^3} e^{ - m^2 s} [
(qEs) f(s) - g(s) ] \nonumber\\&& + i \frac{(2 |\sigma| +
1)(qE)^{2}}{16 \pi^{3}} \sum_{n=1}^{\infty} \frac{(-1)^{(2 |\sigma|
+ 1) (n+1)}}{n^{2}} e^{-\frac{\pi m^{2}n}{qE}}, \label{spin-E3}
\end{eqnarray}
where $f(s)= {\rm cosec}(qEs)$ and $ g(s) = 1 + (qE s)^{2}/6$ for
bosons, and $f(s) = \cot(qE s)$ and $g(s) = 1/s - (qES)^3/3$ for
fermions. Note that the effective action is independent of the
gauges used as expected. The vacuum persistence (twice the imaginary
part) can be written as
\begin{eqnarray}
2 {\rm Im} ({\cal L}^{\rm eff}) = - (-1)^{2 |\sigma| + 1} \frac{(2
|\sigma| + 1) qE}{(2\pi)} \int \frac{d^2 {\bf k}_{\perp}}{(2\pi)^2}
\ln [1 - (-1)^{2 |\sigma| + 1} {\cal N}_{\bf k}],
\end{eqnarray}
where ${\cal N}_{\bf k}$ is the mean number of the produced pairs:
\begin{eqnarray}
{\cal N}_{\bf k} = e^{- \frac{\pi (m^2 + {\bf k}_{\perp}^2)}{qE}}.
\end{eqnarray}

\section{Apparent Spin-Statistics Inversion for the Pair-Production Rate}

In Ref. \cite {Kim-Page07} the pair-production rate by a spatially
localized electric field in the Coulomb gauge, $E(x) = -
dA_0(x)/dx$, is given by
\begin{eqnarray}
{\cal N}_{\omega {\bf k}_{\perp} \sigma} = e^{- {\rm Re} ({\cal
S}_{\omega {\bf k}_{\perp} \sigma})}, \label{WKB pair}
\end{eqnarray}
where the WKB approximation of the action ${\cal S}$ is
\begin{eqnarray}
{\cal S}_{\omega {\bf k}_{\perp} \sigma}^{(0)} = \oint \sqrt{-
Q_{\omega {\bf k}_{\perp} \sigma}(x)} dx, \quad Q_{\omega {\bf
k}_{\perp} \sigma} = m^2 + {\bf k}_{\perp}^2 + 2 i \sigma q E(x) -
(\omega - qA_0 (x))^2.
\end{eqnarray}
Here, the contour integral is taken outside of the branch cut
connecting two roots of $Q_{\omega {\bf k}_{\perp} \sigma}(x)$ in a
complex plane $x$. For the Sauter-type electric field, $E(x) = E_0
{\rm sech}^2 (x/L)$, the pair-production rate (\ref{WKB pair}) turns
out to be a good approximation scheme \cite{Kim-Page07}, whose total
mean number up to quadratic terms of momenta is equivalent to the
worldline instanton plus the prefactor \cite{Dunne-Schubert,DWGS}.

An interesting point is observed that the pair-production rate for
fermions and bosons is better approximated by the WKB action for
bosons and fermions, respectively. This apparent inversion of
spin-statistics can be shown for $E(x) = E_0 {\rm sech}^2 (x/L)$ and
$E(t) = E_0 {\rm sech}^2 (t/T)$ by including the next-to-leading
order correction to the WKB action,
\begin{eqnarray}
{\cal S}_{\omega {\bf k}_{\perp} \sigma}^{(2)} = - i \oint \Bigl(-
\frac{\sqrt{- Q_{\omega {\bf k}_{\perp} \sigma}} [5 (Q_{\omega {\bf
k}_{\perp} \sigma}')^2 - 4 Q_{\omega {\bf k}_{\perp} \sigma}
Q_{\omega {\bf k}_{\perp} \sigma}'']}{32 Q_{\omega {\bf k}_{\perp}
\sigma}^3} \Big) dx.
\end{eqnarray}
In fact, the following relations hold
\begin{eqnarray}
{\rm Re} ({\cal S}_{\rm sp}^{(0)} + {\cal S}_{\rm sp}^{(2)}) \approx
{\cal S}_{\rm sc}^{(0)},
\end{eqnarray}
and
\begin{eqnarray}
{\cal S}_{\rm sc}^{(0)} + {\cal S}_{\rm sc}^{(2)} \approx {\rm Re}
({\cal S}_{\rm sp}^{(0)}).
\end{eqnarray}
The above relation is numerically confirmed for the profile $E(t) =
E_0 {\rm sech}^2 (t/T)$, which can be shown by comparing Figs. 1 and
3, and Figs. 2 and 4, respectively, of Ref. \cite{Dumlu} and for the
profile $E(t) = E_0 \cos (\omega t + \phi) e^{- t^2/(2 \tau^2)}$
\cite{HADG}. The physical reasoning for this apparent inversion of
spin-statistics is not known. The inversion of spin-statistics is
also known for the power spectrum of the vacuum noise seen by a
uniformly accelerated observer in odd dimensions \cite{Takagi}. The
connection of Unruh effect with Schwinger mechanism has been studied
in Ref. \cite{Kim07}. However, the WKB action for Schwinger
mechanism depends on the dimensionality only through the transverse
momenta, while Unruh effect is approximately the Boltzmann
distribution and is independent of spin. To exploit the underlying
physics for this apparent inversion of spin-statistics requires a
further study.

\begin{theacknowledgments}
The author would like to thank H.~K.~Lee, D.~N.~Page and Y.~Yoon for
early collaborations and useful discussions and Prof. S.~P.~Chia for
the warm hospitality during IMFP2009. This work was supported by the
Korea Research Foundation Grant funded by the Korean Government
(MOEHRD) (KRF-2007-C00167). The travel for IMFP2009 was supported by
supported by the Korea Research Council of Fundamental Science and
Technology (KRCF)
\end{theacknowledgments}

\end{document}